\documentstyle[a4,12pt]{article}
\begin{document}
\parskip=5pt plus 1pt minus 1pt

\begin{flushright}
{\bf MPI-PhT/95-91} \\
{September 1995}
\end{flushright}

\vspace{0.2cm}
\begin{center}
{\Large\bf Lepton Mass Hierarchy and Neutrino Oscillations}
\end{center}

\vspace{0.2cm}

\begin{center}
{\bf Harald Fritzsch} \footnote{Supported in part by DFG-contract 412/22-1,
EEC-contract SC1-CT91-0729, and EEC-contract CHRX-CT94-0579 (DG 12 COMA)} \\
{\sl Sektion Physik, Theoretische Physik, Universit$\ddot{a}$t
M$\ddot{u}$nchen,}\\
{\sl Theresienstrasse 37, D-80333 M$\ddot{u}$nchen, Germany} \\
{\sl and} \\
{\sl Max-Planck-Institut f$\sl\ddot{u}$r Physik ---
Werner-Heisenberg-Institut,}\\
{\sl F$\sl\ddot{o}$hringer Ring 6, D-80805 M$\sl\ddot{u}$nchen, Germany}
\end{center}

\begin{center}
{\bf Zhi-zhong Xing} \footnote{Electronic address:
Xing@hep.physik.uni-muenchen.de} \\
{\sl Sektion Physik, Theoretische Physik, Universit$\ddot{a}$t
M$\ddot{u}$nchen,} \\
{\sl Theresienstrasse 37, D-80333 M$\ddot{u}$nchen, Germany}
\end{center}
\vspace{1.6cm}

\begin{abstract}
Starting from the symmetry of lepton flavor democracy, we propose and discuss a
simple
pattern for the mass generation and flavor mixing of the charged leptons and
neutrinos.
The three neutrino masses are nearly degenerate, and the flavor mixing angles
can be calculated. The observed deficit of solar and atmospheric neutrinos
can be interpreted as a consequence of the near degeneracy and large
oscillations of
$\nu_e$, $\nu_{\mu}$ and $\nu_{\tau}$ in the vacuum. Our {\it Ansatz}
can also accommodate the cosmological requirement for hot dark matter and the
current
data on neutrinoless $\beta\beta$-decay.
\end{abstract}

\newpage

In the standard model of the electroweak interactions the masses of quarks and
leptons as well as the flavor mixing angles enter as free parameters. Further
insights
into the yet unknown dynamics of the mass generation require steps beyond the
standard
model. The first one in this direction could be the identification of specific
patterns
and symmetries and the associated symmetry breaking.

\vspace{0.3cm}

Recently a number of authors \cite{Demo,HF1,HF2} have stressed that the
observed
hierarchies in the lepton-quark mass spectrum could be interpreted as a hint
towards
the significance of a ``democratic'' mass matrix both for the up- and down-type
quarks
and for the charged leptons:
$$
M_{0i} \; =\; c_i \left ( \begin{array}{ccc}
1 	& 1	& 1 \\
1	& 1	& 1 \\
1	& 1	& 1
\end{array} \right ) \;
\eqno{(1)}
$$
($i$ stands for $u, d$ in case of quarks and $l$ in case of the charged
leptons).
These mass matrices are supposed to be valid in the limit where the first and
second
family of leptons and quarks are massless. Small violations of the ``democratic
symmetry'' can account for the masses of the second and first family of quarks
as
well as for the flavor mixing angles (see, e.g. refs. \cite{HF1,HF2}).

\vspace{0.3cm}

In this paper we should like to point out that an application of similar ideas
to the
leptons can lead to a surprisingly simple pattern for the mass generation and
flavor
mixing in the neutrino sector. In particular the mixing angles describing
neutrino
oscillations are large, calculable and consistent with experimental
constraints.

\vspace{0.3cm}

In the ``democratic limit'' only the third family of quarks and leptons (i.e.
the
$(t,b)$ system and the $\tau$-lepton) acquire masses. Suppose a mass would also
be
introduced for the $\tau$-neutrino along the same line. In this case we would
obtain
a massive neutrino $\nu^{~}_{\tau}$, which could be either a Majorana or a
Fermi-Dirac
state, and the neutrino mass matrix takes the form:
$$
M_{0 \nu} \; =\; c_{\nu} \left ( \begin{array}{ccc}
1	& 1	& 1 \\
1	& 1	& 1 \\
1	& 1	& 1
\end{array} \right ) \; ,
\eqno{(2)}
$$
($m_{\nu^{~}_{\tau}}=3c_{\nu}$).
Since according to astrophysical constraints the $\nu_{\tau}$-state must be
very
light, i.e. not heavier than about 30 eV, we would have a situation in which
the
constants $c_{\nu}$ and $c_{l}$ for the various flavor channels differ by at
least
eight orders of magnitude ($c_{\nu}/c_{l}< 30~ {\rm eV}/m_{\tau}\sim 10^{-8}$).
We find that such a tiny ratio is very unnatural, and one is invited to look
for
another possibility to introduce the neutrino masses.

\vspace{0.3cm}

In our view the simplest way to avoid the problem mentioned above is to suppose
that the constant $c_{\nu}$ vanishes, i.e. the neutrinos do not receive any
mass
contribution in the ``democratic limit''. We do not attempt to discuss the
dynamical
reason for the vanishing of $c_{\nu}$ except for mentioning that it would
follow
if one could establish a multiplicative relation between the fermion masses in
the
``democratic limit'' and their electric charges, i.e. the vanishing of
$c_{\nu}$
would be directly related to the fact that the neutrinos are electrically
neutral.
If $c_{\nu}$ vanishes, it is automatically implied that there exists a
qualitative
difference between the neutrino sector and the charged lepton sector. In
particular
it is expected that the neutrino masses are small compared to the main entry in
the
charged lepton mass matrix $c_{l}=m_{\tau}/3$, and in particular there would be
no
reason why the hierarchical pattern observed for the charged lepton masses
should
repeat itself for the neutrino masses, i.e. the three neutrino masses could be
of
the same order in magnitude.

\vspace{0.3cm}

In the absence of the ``democratic'' neutrino term, one would have:
$$
M_{l} \; =\; c_{l} \left ( \begin{array}{ccc}
1	& 1	& 1 \\
1	& 1	& 1 \\
1	& 1	& 1
\end{array} \right ) + \Delta M_{l} \; , ~~~~~~~~
M_{\nu} \; =\; 0 + \Delta M_{\nu} \; ,
\eqno{(3)}
$$
where $\Delta M_{l}$ and $\Delta M_{\nu}$ are the symmetry breaking terms for
the
charged leptons and neutrinos, respectively. As discussed previously
\cite{HF3}, a simple
breaking term would be a diagonal mass shift for the ``democratic''
eigenstates, i.e.
$$
\Delta M_{l} \; =\; \left ( \begin{array}{ccc}
\delta_{l}	& 0	& 0 \\
0	& \varrho^{~}_{l}	& 0 \\
0	& 0	& \varepsilon^{~}_{l}
\end{array} \right ) \; , ~~~~~~~~
M_{\nu} \; =\; \left ( \begin{array}{ccc}
\delta_{\nu}	& 0	& 0 \\
0	& \varrho_{\nu}	& 0 \\
0	& 0	& \varepsilon_{\nu}
\end{array} \right ) \; .
\eqno{(4)}
$$
Both $M_{l}$ and $M_{\nu}$ are real matrices, i.e. $CP$ symmetry is preserved
for
the leptons. The neutrino mass matrix is already diagonal (eigenvalues:
$\delta_{\nu}, \varrho_{\nu}, \varepsilon_{\nu}$), while the mass matrix for
the
charged leptons needs to be diagonalized. Apart from small corrections from
$\Delta M_l$, the main effect of the diagonalization is to diagonalize the
``democratic matrix'' $M_{0l}$ by the transformation $UM_{0l}U^{\dagger} =
M^{l}_{\rm H}$, where $M^{l}_{\rm H}$ is the ``hierarchical'' matrix:
$$
M^{l}_{\rm H} \; =\; c_{l} \left ( \begin{array}{ccc}
0	& 0	& 0 \\
0	& 0	& 0 \\
0	& 0	& 3
\end{array} \right ) \; , ~~~~~~~~
U \; =\; \left ( \begin{array}{ccc}
\displaystyle\frac{1}{\sqrt{2}}	& -\displaystyle\frac{1}{\sqrt{2}}	& 0 \\
\displaystyle\frac{1}{\sqrt{6}}	& \displaystyle\frac{1}{\sqrt{6}}	&
-\displaystyle\frac{2}{\sqrt{6}} \\
\displaystyle\frac{1}{\sqrt{3}}	& \displaystyle\frac{1}{\sqrt{3}}	&
\displaystyle\frac{1}{\sqrt{3}}
\end{array} \right ) \; .
\eqno{(5)}
$$
In a good approximation the leptonic flavor mixing matrix is given by the
matrix
$U$ above, i.e. the leptonic doublets are given by
$$
\left ( \begin{array}{ccc}
\displaystyle\frac{1}{\sqrt{2}}(\nu_{1}-\nu_{2})	&
\displaystyle\frac{1}{\sqrt{6}}
(\nu_{1}+\nu_{2}-2\nu_{3})	&
\displaystyle\frac{1}{\sqrt{3}}(\nu_{1}+\nu_{2}+\nu_{3}) \\
e^{-}	& \mu^{-}	& \tau^{-}
\end{array} \right ) \; ,
\eqno{(6)}
$$
where $\nu_{1}, \nu_{2}, \nu_{3}$ are the neutrino mass eigenstates.

\vspace{0.3cm}

In order to estimate the corrections from the symmetry breaking term $\Delta
M_l$,
we consider an illustrative example, in which $\delta_{l}=-\varrho^{~}_{l}$ is
taken.
This case is of particular interest. The mass matrix in the hierarchical basis,
i.e. the matrix $M_{\rm H}^l$, has a vanishing (1,1) element, and the mixing
angles
can be completely expressed in terms of mass eigenvalues. Taking into account
the
higher order terms from the symmetry breaking, one finds for the lepton mixing
matrix $V$:
$$
V \; = \; U + \sqrt{\frac{m_{e}}{m_{\mu}}} \left (
\begin{array}{ccc}
\displaystyle\frac{1}{\sqrt{6}}	& \displaystyle\frac{1}{\sqrt{6}}	&
-\displaystyle\frac{2}{\sqrt{6}} \\
-\displaystyle\frac{1}{\sqrt{2}}	& \displaystyle\frac{1}{\sqrt{2}}	& 0 \\
0	& 0	& 0
\end{array} \right ) \; +\; \frac{m_{\mu}}{m_{\tau}} \left (
\begin{array}{ccc}
0	& 0	& 0 \\
-\displaystyle\frac{1}{\sqrt{6}}	& -\displaystyle\frac{1}{\sqrt{6}}	&
-\displaystyle\frac{1}{\sqrt{6}} \\
\displaystyle\frac{1}{2\sqrt{3}}	& \displaystyle\frac{1}{2\sqrt{3}}	&
-\displaystyle\frac{1}{\sqrt{3}}
\end{array} \right ) \; .
\eqno{(7)}
$$
Here $\sqrt{m_{e}/m_{\mu}}\approx 0.0696$ and $m_{\mu}/m_{\tau}\approx 0.0594$
\cite{PDG}.
In general the real mixing matrix $V$ can be parametrized in terms of three
Euler
angles, which we define in analogy to the quark mixing angles \cite{PDG}:
$$
V \; =\; \left ( \begin{array}{ccc}
c_{12}c_{13}	& s_{12}c_{13}	& s_{13} \\
-s_{12}c_{23}-c_{12}s_{23}s_{13}	& c_{12}c_{23}-s_{12}s_{23}s_{13}	&
s_{23}c_{13} \\
s_{12}s_{23}-c_{12}c_{23}s_{13}	& -c_{12}s_{23}-s_{12}c_{23}s_{13}	&
c_{23}c_{13}
\end{array} \right ) \;
\eqno{(8)}
$$
($c_{ij}=\cos\theta_{ij}$, $s_{ij}=\sin\theta_{ij}$). Comparing
eqs. (7) and (8), the three mixing angles can be determined as follows:
$$
\tan\theta_{12} =  -1+\frac{2}{\sqrt{3}}\sqrt{\frac{m_{e}}{m_{\mu}}} \; ,
{}~~~~~~
\tan\theta_{23} =  -\sqrt{2}-\frac{3}{\sqrt{2}}\frac{m_{\mu}}{m_{\tau}} \; ,
{}~~~~~~
\tan\theta_{13} =  -\frac{2}{\sqrt{6}}\sqrt{\frac{m_{e}}{m_{\mu}}} \; .
\eqno{(9)}
$$
The angle $\theta_{13}$ is very small, compared to $\theta_{12}$ and
$\theta_{23}$,
and vanishes in the limit $m_e \rightarrow 0$. In this special case one obtains
a
two-angle parametrization of $V$.

\vspace{0.3cm}

In the more general case, where $\delta_l \neq -\varrho^{~}_l$, the mixing
angles
cannot be calculated only in terms of the mass eigenvalues. Nevertheless, the
pattern
that $V$ is very close to $U$ persists. In the following we shall use
$V=U$ to discuss neutrino oscillations and the associated problems. We
emphasize that
in this limit the mixing angles are algebraic numbers independent of the lepton
masses.
The three angles can all be arranged to be in the second quadrant:
$\theta_{12} = 135^{0}$, $\theta_{23} = 125.3^{0}$ and $\theta_{13} = 180^{0}$.
In
particular the electron neutrino $\nu_{e}$ is a mixture of the two mass
eigenstates
$\nu_{1}$ and $\nu_{2}$, with maximal mixing between the two. Both $\nu_{\mu}$
and
$\nu_{\tau}$ are composed of all three mass eigenstates (see eq. (6)). It is
useful
to compare the structure of the neutrino states with the $SU(3)$ wave functions
of
the pseudoscalar mesons in the chiral limit of $SU(3)_{L}\times SU(3)_{R}$
symmetry \cite{HF3}. In terms of $(\bar{q}q)$ states ($q$: quark field), one
has:
$$
\pi^{0} \; =\; \frac{1}{\sqrt{2}}|\bar{u}u -\bar{d}d\rangle \; , ~~~~~
\eta \; =\; \frac{1}{\sqrt{6}}|\bar{u}u +\bar{d}d -2\bar{s}s\rangle \; , ~~~~~
\eta^{'} \; =\; \frac{1}{\sqrt{3}}|\bar{u}u +\bar{d}d +\bar{s}s\rangle \; .
\eqno{(10)}
$$
Thus the flavor eigenstates $\nu_{e}$, $\nu_{\mu}$ and $\nu_{\tau}$ correspond
to
$\pi^{0}$, $\eta$ and $\eta^{'}$, and the neutrino mixing angles are identical
to
the mixing angles of the pseudoscalar mesons in the chiral limit.

\vspace{0.3cm}

Analogous to the quark fields, the lepton fields enter the weak charged current
in a
mixed form described by $V$ \cite{HF4}. Neutrino beams will oscillate. Assuming
a $\nu_{i}$
to have been produced at proper time $t=0$, with momentum {\bf P} ($|{\bf P}|
>> m_{i}$,
$i=1,2,3$), then the probability for observing a $\nu_{l}^{~}$ generating from
$\nu_{i}$
at time $t$ is given by
$$
P_{li} \; =\; \sum^{3}_{k=1}\left (V_{ik}V_{lk}\right )^{2} + 2\sum_{k>j} \left
\{ \left (
V_{ij}V_{lj}V_{ik}V_{lk}\right ) \cos\left [ \left (E_{k}-E_{j}\right )t \right
] \right \} \; .
\eqno{(11)}
$$
In the approximation of $V = U$, we obtain:
$$
P_{\mu e} \; = \; \frac{1}{6} - \frac{1}{6}\cos \left [(E_{2}-E_{1})t \right ]
\; , ~~~~~~
P_{\tau e} \; =\; \frac{1}{3} - \frac{1}{3}\cos \left [(E_{2}-E_{1})t \right ]
\; ,
\eqno{(12{\rm a})}
$$
and
$$
P_{\tau\mu} \; =\; \frac{1}{3} + \frac{1}{9}\cos \left [(E_{2}-E_{1})t \right ]
-\frac{2}{9}\cos \left [(E_{3}-E_{1})t \right ] - \frac{2}{9}\cos \left
[(E_{3}-E_{2})t \right ] \; .
\eqno{(12{\rm b})}
$$
By use of $E_{i}\approx |{\bf P}|+m^{2}_{i}/(2|{\bf P}|)$, the above
transition probabilities can be expressed in terms of the quantities
$F_{ij}=1.27\Delta m^{2}_{ij}L/|{\bf P}|$, where $\Delta m^{2}_{ij} =
m^{2}_{i}-m^{2}_{j}$ (in unit of eV$^{2}$) denotes the mass-squared
difference of neutrinos and $L$ (in unit of km/GeV or m/MeV) is the
distance from the neutrino's production point to its interaction point.
One obtains:
$$
P_{\mu e} \; = \frac{1}{3}\sin^{2}F_{21} \; , \;\;\;\;\;\;\;\;
P_{\tau e} \; = \; \frac{2}{3}\sin^{2}F_{21} \; ,
\eqno{(13{\rm a})}
$$
and
$$
P_{\tau\mu} \; = \; -\frac{2}{9}\sin^{2}F_{21}
+ \frac{4}{9}\sin^{2}F_{31} + \frac{4}{9}\sin^{2}F_{32} \; .
\eqno{(13{\rm b})}
$$
Note that $P_{li}=P_{il}$ is a consequence of $CP$ symmetry, and the unitarity
of $U$
implies $P_{ei}+P_{\mu i}+P_{\tau i} =1$ for each $\nu_{i}^{~}$ ($i=e$, $\mu$
or $\tau$).

\vspace{0.3cm}

Let us confront the above results with the solar and atmospheric neutrino
experiments.
The experimental data of neutrino oscillations are usually presented as allowed
regions on a $\Delta m^{2}_{ij} - \sin^{2}\Theta_{ij}$ plot in the assumption
of
two-flavor oscillations, although there are (at least) three flavors of
neutrinos.
Here the mixing angles $\Theta_{ij}$ ($i,j = 1,2,3$), which can be derived from
$\theta_{ij}$ in the lepton mixing matrix (8), do measure the magnitudes of
neutrino
oscillations. The current evidence for the deficit of solar neutrinos mainly
comes
from four experiments \cite{Solar}. Such a deficiency can be interpreted as a
$\nu_{e}$-neutrino disappearance experiment due to its oscillations to
$\nu_{\mu}$ and
$\nu_{\tau}$ neutrinos. From eq. (13) we obtain
$$
P_{ee} \; = \; 1- \sin^{2}F_{21} \; ,
\eqno{(14)}
$$
which incorporates the maximal mixing $\sin^{2}2\Theta_{21} = 1$. Analyses of
solar
neutrino data on the basis of two-neutrino oscillations in the vacuum have
found a small
but stable parameter space for the mixing angle and mass-squared difference
\cite{Pet}:
$\sin^{2}2\Theta_{21} = 0.59...1.0$ and $\Delta m^{2}_{21} = (0.47...9.8)\times
10^{-11}$ eV$^{2}$, with respect to changes of the total fluxes of $^8$B and
$^7$Be
neutrinos. This implies that our model favors the solution of long-wavelength
vacuum oscillations to the solar neutrino problem. In addition, $\Delta
m^{2}_{21}\sim
10^{-11}$ eV$^{2}$ implies that the neutrino mass eigenstates $\nu_{1}$ and
$\nu_{2}$
are essentially degenerate.

\vspace{0.3cm}

The atmospheric neutrinos originate from cosmic ray showers in the upper
atmosphere,
in particular from pion and subsequent muon decays. Depletion of $\nu_{\mu}$
relative
to $\nu_{e}$ has been observed in several experiments \cite{Atmos}. In the
assumption that
this problem is solved by $\nu_{\mu}-\nu_{\tau}$ oscillations, the combined
data
indicate $\Delta m^{2}_{32}\sim 10^{-2}$ eV$^{2}$ and $\sin^{2}2\Theta_{32} =
0.4...1.0$.
{}From eq. (13) one obtains
$$
P_{\mu \mu} \; = \; 1- \frac{1}{9}\sin^{2}F_{21} - \frac{4}{9}\sin^{2}F_{31}
- \frac{4}{9}\sin^{2}F_{32} \; .
\eqno{(15{\rm a})}
$$
Since we have $L\leq 1.3\times 10^{4}$ km and $|{\bf P}|\geq 0.2$ GeV,
$\Delta m^{2}_{21}\sim 10^{-11}$ eV$^{2}$ obtained above implies that the
$\sin^{2}F_{21}$
term in eq. (15a) can be safely neglected. Note that the extreme smallness of
$\Delta m^{2}_{21}$ leads to $\Delta m^{2}_{32}$ $\approx$ $\Delta m^{2}_{31}$
(i.e. $F_{32}\approx F_{31}$), and then we obtain:
$$
P_{\mu \mu} \; \approx \; 1- \frac{8}{9}\sin^{2}F_{32} \; ,
\eqno{(15{\rm b})}
$$
i.e. the deficit of $\nu_{\mu}$ mainly arises from $\nu_{\mu}-\nu_{\tau}$
oscillations.
This result corresponds to the mixing magnitude $\sin^{2}2\Theta_{32}\approx
8/9$, which is
consistent with the experimental constraints. Accordingly we have
$\Delta m^{2}_{32}\sim 10^{-2}$ eV$^{2}$, as indicated by the allowed region on
the $\Delta m^{2}_{32} - \sin^{2}\Theta_{32}$ plot \cite{Atmos}.

\vspace{0.3cm}

Let $N_{e}$ and $N_{\mu}$ be the original $\nu^{~}_{e}$ and $\nu^{~}_{\mu}$
fluxes
respectively, at the point of production somewhere in the atmosphere. Due to
oscillations
after travelling a distance, the effective fluxes $\hat{N}_{e}$ and
$\hat{N}_{\mu}$ at
the point of detection are given as
$$
\hat{N}_{e} \; = \; N_{e} \displaystyle\left (
P_{ee}+\frac{N_{\mu}}{N_{e}}P_{e\mu}\right )  \;
= \; N_{e} \displaystyle\left [ 1-\left
(1-\frac{1}{3}\frac{N_{\mu}}{N_{e}}\right )
\sin^{2}F_{21} \right ] \;
\eqno{(16{\rm a})}
$$
and
$$
\begin{array}{lll}
\hat{N}_{\mu} & = & N_{\mu} \displaystyle\left (
P_{\mu\mu}+\frac{N_{e}}{N_{\mu}}P_{\mu e}\right ) \\
 & = & N_{\mu} \displaystyle\left [ 1-\frac{1}{3}\left
(\frac{1}{3}-\frac{N_{e}}{N_{\mu}}\right )
\sin^{2}F_{21} - \frac{4}{9}\sin^{2}F_{31} - \frac{4}{9}\sin^{2}F_{32} \right ]
\; .
\end{array}
\eqno{(16{\rm b})}
$$
Here $N_{\mu}/N_{e}\approx 2$, obtained from the Monte Carlo calculations for
low-energy
$\nu^{~}_{\mu}$ and $\nu^{~}_{e}$ fluxes \cite{MC}. To the approximation made
in eq. (15b),
we obtain $\hat{N}_{e}\approx N_{e}$ and
$$
\frac{\hat{N}_{\mu}/N_{\mu}}{\hat{N}_{e}/N_{e}} \; \approx \;
1-\frac{8}{9}\sin^{2}F_{32} \; ,
\eqno{(17)}
$$
which can be confronted with the experimental data \cite{Atmos}.

\vspace{0.3cm}

In our approach we find that the neutrino mass eigenstates $\nu_{1}$ and
$\nu_{2}$
are essentially degenerate. Furthermore, due to the constraint
$\Delta m^{2}_{32}\approx \Delta m^{2}_{31}\sim 10^{-2}$ eV$^{2}$, all three
neutrino states must be nearly degenerate. If one identifies the dark matter of
the universe (or at least its hot dark matter component) with neutrino matter,
we must have $m_{1}+m_{2}+m_{3} \approx 3m_{1}\approx 7...25$ eV. For example,
if $m_{i}\approx 2.5$ eV, one has $m_{1}+m_{2}+m_{3} \approx 7.5$ eV. Recently
a related case has been discussed in ref. \cite{Moha}.

\vspace{0.3cm}

Within our approach, the degeneracy of three neutrino
masses is suggestive that the mass generation of neutrinos proceeds in three
steps.
At the first step a diagonal universal mass is introduced:
$|\varepsilon_{\nu}|=
|\varrho_{\nu}|=|\delta_{\nu}|$. Secondly, the degeneracy between $\nu_{3}$ and
$\nu_{2}$ is lifted: $|\varepsilon_{\nu}|\neq |\varrho_{\nu}|=|\delta_{\nu}|$.
Finally,
a tiny difference between $|\varrho_{\nu}|$ and $|\delta_{\nu}|$ appears.

\vspace{0.3cm}

We mentioned before that the neutrino mass terms could be either of Fermi-Dirac
type
or of Majorana type. In our approach the neutrino mass terms correspond to the
small
breaking terms in the charged lepton sector, i.e. they are expected to be
small; and
it does not seem necessary for us to consider special mechanisms like the
so-called
``see-saw'' mechanism \cite{SS} for Majorana states to suppress the neutrino
masses.
Thus we see no particular strong reason why the neutrino mass terms should be
of
Majorana type.

\vspace{0.3cm}

In the case of Majorana masses, the difficulty arises to fulfill the bound
$\langle
m \rangle \leq 0.7$ eV for neutrinoless $\beta\beta$-decay \cite{Beta}, where
$\langle m \rangle$ is an effective mass factor. It is known that the
$(\beta\beta)_{0\nu}$-decay amplitude depends on the masses of Majorana
neutrinos
$m_{i}$ and on the elements of the lepton mixing matrix $V_{ek}$:
$$
\langle m \rangle \; \sim \;
\sum^{3}_{k=1}\left [ V_{ek}^{2}m_{k}\lambda (\psi_{k}) \right ] \; ,
\eqno{(18)}
$$
where $\lambda(\psi_{k})$ is the $CP$ parity of the Majorana field $\psi_{k}$.
If $\lambda(\psi_{1})=+1$ and $\lambda(\psi_{2})=-1$ (or vice versa), one finds
that $\langle m \rangle \sim (m_{1}-m_{2})/2$, which is considerably suppressed
due to the nearly degeneracy of $m_{1}$ and $m_{2}$. This implies that there
may
exist two Majorana neutrinos with opposite $CP$ eigenvalues, and their relative
$CP$ parities are in principle observable in $(\beta\beta)_{0\nu}$-decay
\cite{Wolf}.
Within our approach this possibility exists. Thus a degenerate Majorana mass of
about 2.5 eV for all three neutrinos need not be in conflict with the data on
neutrinoless $\beta\beta$-decay.

\vspace{0.3cm}

It is of interest to note that the required
cancellation in $\langle m\rangle$ takes place in the specific case
$\varrho_{\nu}\approx -\delta_{\nu}$, parallel to the condition $\varrho^{~}_l
=
- \delta_l$ taken above. In this sense, one is invited to speculate about a
similarity of the perturbative structures of the charged lepton and neutrino
mass matrices, although they are obviously very different in the ``democratic
limit''.

\vspace{0.3cm}

In our model for the neutrino masses the mismatch between the ``democratic''
mass
matrix of the charged leptons and the neutrino mass matrix, where the
``democratic''
mass term is absent, generates large effects of flavor mixing. The associated
mixing angles are large, calculable and in lowest order independent of the
lepton
masses. The neutrino masses consistent with the experimental constraints
are nearly degenerate. One may interpret the hot dark matter component of
the cosmic mass density as neutrino matter, in which the three neutrinos are
expected to have masses of the order of a few eV. If our description of the
neutrino
mass and mixing pattern is correct, we expect that a positive signal for
neutrino
oscillations will be obtained by investigating a $\nu_{\mu}$-beam at distances
of
the order of several hundred kilometers from its production point \cite{Exp}.
In our
view it would be of high interest to carry out such long base-line experiments.

\vspace{0.5cm}

We are indebted to Profs. P. Minkowski and S. Petcov for useful discussions.
The
work of Z.X. was supported by the Alexander von Humboldt Foundation.



\vspace{0.6cm}

\begin{thebibliography}{99}
\bibitem{Demo} H. Harari, H. Haut, and J. Weyers, Phys. Lett. B78, 459 (1978);
Y. Chikashige, G. Gelmini, R.P. Peccei, and M. Roncadelli, Phys. Lett. B94, 499
(1980);
H. Fritzsch, in: Proc. of Europhys. Conf. on Flavor Mixing in Weak
Interactions, Erice, Italy (1984);
C. Jarlskog, in: Proc. of Int. Symp. on Production and Decay of Heavy Flavors,
Heidelberg,
Germany (1986);
P. Kaus and S. Meshkov, Mod. Phys. Lett. A3, 1251 (1988);
Y. Koide, Phys. Rev. D39, 1391 (1989);
M. Tanimoto, Phys. Rev. D41, 1586 (1990);
G.C. Branco, J.I. Silva-Marcos, and M.N. Rebelo, Phys. Lett. B237, 446 (1990).

\bibitem{HF1} H. Fritzsch and J. Plankl, Phys. Lett. B237, 451 (1990);
H. Fritzsch, Phys. Lett. B289, 92 (1992);
H. Fritzsch and Z.Z. Xing, Phys. Lett. B353, 114 (1995).

\bibitem{HF2} H. Fritzsch and D. Holtmannsp$\rm\ddot{o}$tter, Phys. Lett. B338,
290 (1994).

\bibitem{HF3} H. Fritzsch, in: Proc. of XXIXth Rencontres de Moriond on
Electroweak
Interactions and Unified Theories, M$\rm\acute{e}$ribel, France (1994);
D. Holtmannsp$\rm\ddot{o}$tter, {\it Diploma Thesis}, University of Munich
(1993);
Y. Koide, Z. Phys. C45, 39 (1989).

\bibitem{PDG} Particle Data Group, L. Montanet et al., Phys. Rev. D50, 1173
(1994).

\bibitem{HF4} H. Fritzsch and P. Minkowski, Phys. Lett. B62, 72 (1976);
S.M. Bilenky and B. Pontecorvo, Phys. Lett. B61, 248 (1976);
S. Eliezer and A.R. Swift, Nucl. Phys. B105, 45 (1976);
For a recent review, see: R.E. Shrock, Phys. Rev. D50, 1385 (1994).

\bibitem{Solar} B.T. Cleveland et al., Proc. XVI Int. Conf. on Neutrino Physics
and Astrophysics,
Nucl. Phys. B (Proc. Suppl.) 38, 47 (1995);
T. Kajita, ICRR-Report 332-94-27 (1994);
P. Anselmann et al., GALLEX Collaboration, LNGS 95/37 (1995);
J.N. Abdurashitov et al., Nucl. Phys. B (Proc. Suppl.) 38, 60 (1995).

\bibitem{Pet} P.I. Krastev and S.T. Petcov, preprint SISSA 9/95/EP (1995);
Phys. Rev. Lett. 72, 1960 (1994);
Phys. Lett. B299, 99 (1993);
Z.G. Berezhiani and A. Rossi, Phys. Rev. D51, 5229 (1995);
V. Barger, R.J.N. Phillips, and K. Whisnant, Phys. Rev. Lett. 69, 3135 (1992);
A. Acker, S. Pakvasa, and J. Pantaleone, Phys. Rev. D43, 1754 (1991).

\bibitem{Atmos} K.S. Hirata et al., Phys. Lett. B280, 146 (1992);
R. Becker-Szendy et al., Phys. Rev. D46, 3720 (1992);
P.J. Litchfield, in: Proc. of Int. Europhys. Conf. on High Energy Physics,
Marseille, France (1993).

\bibitem{MC} G. Barr et al., Phys. Rev. D39, 3532 (1989); Phys. Lett. B214, 147
(1988).

\bibitem{Moha} D.O. Caldwell and R.N. Mohapatra, Phys. Rev. D48, 3259 (1993);
R.N. Mohapatra and S. Nussinov, Phys. Lett. B346, 75 (1995).

\bibitem{SS} M. Gell-Mann, P. Ramond, and R. Slansky, in: {\it Supergravity},
ed. F. van Nieuwenhuizen
and D. Freedman (North Holland, Amsterdam, 1979), p. 315;
T. Yanagida, in: Proc. of Workshop on Unified Theory and Baryon Number of the
Universe, KEK, Japan, 1979;
S. Weinberg, Phys. Rev. Lett. 43, 1566 (1979).

\bibitem{Beta} H.V. Klapdor-Kleingrothaus et al., Prog. Part. Nucl. Phys. 32,
261 (1994);
and references therein.

\bibitem{Wolf} L. Wolfenstein, Phys. Lett. B107, 77 (1981);
S.M. Bilenky and S.T. Petcov, Rev. Mod. Phys. 59, 671 (1987);
B. Kayser, in {\it Neutrino Physics}, edited by K. Winter (Cambridge University
Press, 1991), p. 115.

\bibitem{Exp} K. Winter, invited talk presented at the meeting on {\it Neutrino
Astronomy}
at The Royal Society, London, June 1993; Nucl. Phys. B (Proc. Suppl.) 38, 349
(1995);
P. Langacker, invited talk presented at {\it Beyond the Standard Model IV},
Lake Tahoe,
December 1994;
P.F. Harrison, D.H. Perkins, and W.G. Scott, Phys. Lett. B349, 137 (1995).

\end{thebibliography}
\end{document}